\documentclass[a4paper,12pt]{article}
\pdfoutput=1
\usepackage{graphicx,subfigure,amsmath,amssymb}
\usepackage{cite}
\usepackage{color}

\newlength{\dinwidth}
\newlength{\dinmargin}
\begin{document}
\title{\bf  Study of semileptonic $\bar{B}^* \to P \ell \bar{\nu}_\ell$  decays}
\author{Qin Chang$^{a,b}$, Jie Zhu$^{a}$, Xiao-Lin Wang$^{a}$, Jun-Feng Sun$^{a}$ and Yue-Ling Yang$^{a}$\\
{ $^a$\small Institute of Particle and Nuclear Physics, Henan Normal University, Henan 453007, P.~R. China}\\
{ $^b$\small State Key Laboratory of Theoretical Physics, Institute of Theoretical Physics,}\\[-0.2cm]
{     \small Chinese Academy of Sciences, P.~R. China}}
\date{}
 \maketitle

\begin{abstract}
In anticipation of abundant $B^*$ data samples at high-luminosity heavy-flavor experiments in the future,  the tree-dominated semileptonic $\bar{B}^*_{u,d,s} \to P \ell^- \bar{\nu}_\ell$ $(P=D\,,D_s\,,\pi\,,K)$ decays are studied within the Standard Model. After a detailed calculation of the  helicity amplitudes, the theoretical predictions for branching fraction~(decay rate),  lepton spin asymmetry, forward-backward asymmetry and ratio $R_D^{\ast(L)}$ are firstly presented. It is found that the CKM-favored $\bar{B}^* \to D \ell^- \bar{\nu}_\ell$ decays have  relatively large branching fractions of ${\cal O}(10^{-9})$$\sim$${\cal O}(10^{-7})$, and are in the scope of running  LHC and forthcoming SuperKEKB/Belle-II experiments. 
\end{abstract}


\newpage
\section{Introduction}
The semileptonic $B$ meson decays induced by the tree-level $b\to p \ell \bar{\nu}_{\ell}$~($p=u\,,c$) transition provide an ideal ground for testing the Standard Model~(SM) and probing possible hints of new physics~(NP). For instance, (i)  such decays offer ways of extracting the magnitudes of the CKM matrix element $V_{cb}$ and $V_{ub}$. Moreover, the extractions from exclusive vs. inclusive semileptonic decays exhibit a long-standing $\sim 2.5\sigma$ discrepancy~\cite{HFAG,Charles:2004jd}; (ii) The measurements of ratios $R_{D^{(*)}}\equiv \frac{\mathcal{B}(\bar{B}\to D^{(*)}\tau^- \bar{\nu}_\tau)}{\mathcal{B}(\bar{B}\to D^{(*)} \ell^{\prime-} \bar{\nu}_{\ell^{\prime}})}$~$(\ell^{\prime}=\mu\,,e)$ reported by BaBar~\cite{Lees:2012xj,Lees:2013uzd}, Belle~\cite{Huschle:2015rga,Kuhr:2015muu,Abdesselam:2016cgx} and LHCb~\cite{Aaij:2015yra} collaborations exhibit significant deviations from the SM expectations  at $>3\sigma$ level~\cite{Fajfer:2012vx,Lattice:2015rga,Na:2015kha,Fan:2015kna,Fan:2014}, which are the so-called ``$R_{D^{(*)}}$ puzzles''. A lot of efforts have been made for possible solutions within various NP models, for instance, new four fermion operators, two-Higgs-doublet models,  R-parity violating supersymmetry models, leptoquark models, Alternative Left-Right Symmetric Model and so on~\cite{Fajfer:2012jt,Sakaki:2012ft,Datta:2012qk,Bailey:2012jg,Becirevic:2012jf,Tanaka:2012nw, Freytsis:2015qca, Bhattacharya:2015ida,Celis:2012dk,Ko:2012sv,Crivellin:2012ye,Deshpande:2012rr,Sakaki:2013bfa,Greljo:2015mma,Dorsner:2013tla,Biancofiore:2013ki,Bauer:2015knc,Fajfer:2015ycq,Hati:2016thk,Zhu:2016xdg,Alonso:2016gym}. In addition to $B$ mesons, some other hadrons, such as $\Lambda_b$ and $B^*$,  could also decay through $b\to c \ell \bar{\nu}_{\ell}$ transition at quark level, and  therefore, these decay modes would play a similar role as semileptonic $B$ decays mentioned above.

The $\bar{B}^*_q$ meson with quantum number of $n^{2s+1}L_J=1^3S_1$ and $J^P=1^-$ is the partner of $B$ meson in the heavy-meson doublet of $(b\bar{q})$ system~\cite{Isgur:1991wq,Godfrey:1986wj,Eichten:1993ub,Ebert:1997nk}. Its decay occurs mainly through the electromagnetic process $\bar{B}^*_q\to \bar{B}_q\gamma$, and the weak decay modes are generally very rare. Until now, there is no available experimental information for $\bar{B}^*_q$ weak decays due to the limited center-of-mass energy and  integrated luminosity in the previous experiments of heavy flavor physics.
Fortunately, such situation is expected to be improved by the upcoming SuperKEKB/Belle-II experiment~\cite{Abe:2010gxa}, which has started test operations and succeeded in circulating and storing beams in the electron and positron rings recently. For instance, using the target annual integrated luminosity $13\,ab^{-1}/{\rm year}$~\cite{Abe:2010gxa}, the cross section of $\Upsilon(5S)$ production $\sigma(e^+e^-\to\Upsilon(5S))=0.301\,{\rm nb}$~\cite{Huang:2006mf} and the branching fractions of $\Upsilon(5S)$ decays into $B^*$ final states~\cite{PDG14}, one can find that about  $\sim4\times10^9\,(B^{*}_{u,d}+\bar{B}^{*}_{u,d})$ and $\sim2\times10^9\,(B^{*}_s+\bar{B}^{*}_s)$ samples could be collected per year, which implies that the $B^{*}$ decays with branching fractions $> {\cal O}(10^{-9})$ are possible to be observed by Belle-II.

In addition, the running LHC may also provide some experimental information for $B^*$ decays, such as  $B^*_s\to l^+l^-$ decay analyzed in Ref.~\cite{Grinstein:2015aua}, due to the much large beauty production cross section of $pp$ collision compared with $e^+e^-$ collision~\cite{Aaij:2010gn,Bediaga:2012py,Aaij:2014jba}.
Thanks to the rapid development of heavy flavor physics experiments,  the theoretical studies of $B^*$ weak decays, which could provide some useful suggestions and references for the measurements, are urgently required. Recently, a few theoretical evaluations of $B^*$ weak decays have been done, for instance, the studies of the semileptonic $B^*_c$ decays within the QCD sum rules~\cite{Wang:2012hu,Zeynali:2014wya,Bashiry:2014qia}, the pure leptonic $\bar{B}_s^*\to \ell\ell$ and $\bar{B}_{u,c}^*\to \ell \bar{\nu}_\ell$  decays~\cite{Grinstein:2015aua}, the impact of  $\bar{B}_{s,d}^*\to \mu\mu$ on  $\bar{B}_{s,d}\to \mu\mu$ decays~\cite{Xu:2015eev}, and the nonleptonic $\bar{B}^{*0}_{d,s}\to D_{d,s}^+M^-$  ($M=\pi\,,K\,,\rho$ and $K^*$) decays~\cite{Chang:2015jla,Chang:2015ead}. In this paper, we will pay attention to the charged $b\to (u,c)\ell \bar{\nu}_{\ell}$ transitions  induced $\bar{B}^*_{u,d,s}\to P \ell \bar{\nu}_{\ell}$~$(P=D\,,D_s\,,\pi\,,K)$ decays within the SM. Especially, the $\bar{B}^*\to D \ell \bar{\nu}_{\ell}$ decays are suppressed neither by CKM factors (compared to other $\bar{B}^*$ decays) nor by loop factors, and thus expected to be observed with relatively large branching fractions.

Our paper is organized as follows. In section 2, the theoretical framework and calculations of $\bar{B}^*\to P \ell \bar{\nu}_{\ell}$ decays are presented in detail. Section 3 is devoted to the numerical results and discussion. Finally, we give our conclusions in section 4.

\section{Theoretical Framework and Calculation}
\subsection{Effective Hamiltonian and Amplitude}
Within the SM, the quark-level $b \to p \ell^- \bar{\nu}_\ell$ ($p=u\,,c$ and $\ell=\tau\,,\mu\,,e$) transitions occur through $W$-exchange and could be described by the effective low-scale ${\cal O}(m_b)$ Hamiltonian
\begin{eqnarray}\label{eq:heff}
\mathcal{H}_{\rm eff}(b \to p \ell^- \bar{\nu}_\ell)=
\frac{G_F}{\sqrt{2}}\sum_{p=u\,,c}V_{pb}\sum_{\ell=\tau\,,\mu\,,e}
[\bar{p}\gamma_{\mu}(1-\gamma_5)b][\bar{\ell}\gamma^{\mu}(1-\gamma_5)\nu_{\ell}]\,,
\end{eqnarray}
where $G_F$ is Fermi coupling constant, and $V_{pb}$ denotes the CKM matrix elements.
With Eq.~\eqref{eq:heff}, the square matrix element for $\bar{B}^* \to P \ell^- \bar{\nu}_\ell$ decay can be written as
\begin{eqnarray}\label{eq:M2}
|{\cal M}(\bar{B}^* \to P \ell^- \bar{\nu}_\ell)|^2=\frac{G_F^2|V_{pb}|}{2}
|\langle P|\bar{p}\gamma_{\mu}(1-\gamma_5)b|\bar{B}^*\rangle\,
\bar{\ell}\gamma^{\mu}(1-\gamma_5)\nu_{\ell}|^2 \equiv  \frac{G_F^2|V_{pb}|}{2} L_{\mu\nu}H^{\mu\nu}\,,
\end{eqnarray}
in which, leptonic ($L_{\mu\nu}$) and hadronic ($H^{\mu\nu}$) tensors are built from the respective products of the lepton and hadron currents.

Following the strategy for evaluating $B \to D^{*} \ell^- \bar{\nu}_\ell$ decays~\cite{Korner:1987kd,Korner:1989qb,Hagiwara:1989cu,Hagiwara:1989gza}, Eq.~\eqref{eq:M2} can be further expressed as
\begin{eqnarray}\label{eq:M2LI}
|{\cal M}(\bar{B}^* \to P \ell^- \bar{\nu}_\ell)|^2= \frac{G_F^2|V_{pb}|}{2}\sum_{m,m^{\prime},n,n^{\prime}} L(m,n)H(m^{\prime},n^{\prime})g_{mm^{\prime}}g_{nn^{\prime}}\,
\end{eqnarray}
by inserting the completeness relation
\begin{eqnarray}
\sum_{m,n}\bar{\epsilon}_{\mu}(m) \bar	{\epsilon}_{\nu}^*(n)g_{mn}=g_{\mu\nu}\,,
\end{eqnarray}
where $\bar{\epsilon}_{\mu}(\pm,0,t)$ are polarization vectors of virtual $W^*$ boson, $g_{mn}={\rm diag}(+1,-1,-1,-1)$. One may note the point that the quantities $L(m,n)\equiv L^{\mu\nu}\bar{\epsilon}_{\mu}(m)\bar{\epsilon}^*_{\nu}(n)$ and $H(m,n)\equiv H^{\mu\nu}\bar{\epsilon}^*_{\mu}(m)\bar{\epsilon}_{\nu}(n)$ in Eq.~\eqref{eq:M2LI} are Lorentz invariant, and therefore can be evaluated in different reference frames. For convenience of evaluation,  $H(m,n)$ and $L(m,n)$ will be calculated  in the $B^*$-meson rest frame and the virtual $W^*$ rest frame~(or $\ell-\bar{\nu}_\ell$ center-of-mass frame), respectively.

\subsection{Kinematics for $\bar{B}^* \to P \ell^- \bar{\nu}_\ell$ Decays}
Before the further evaluation, we would like to clarify some conventions and definitions for kinematics of $\bar{B}^* \to P \ell^- \bar{\nu}_\ell$ decays used in this paper.

In the rest frame of $B^*$-meson with daughter $P$-meson moving in the positive $z$-direction, the momenta of particles $B^*$, $P$ and virtual $W^*$ could be  written respectively as
 \begin{eqnarray}
 p_{B^*}^{\mu}=(m_{B^*},0,0,0)\,,\quad  p_{P}^{\mu}=(E_P,0,0,|\vec{p}|)\,,\quad q^{\mu}=(q^0,0,0,-|\vec{p}|)\,,
\end{eqnarray}
where $q^0=m_{B^*}-E_P=(m_{B^*}^2-m_P^2+q^2)/2m_{B^*}$ and $|\vec{p}|=\lambda^{1/2}(m_{B^*}^2,m_P^2,q^2)/2m_{B^*}$ with function $\lambda(a,b,c)=a^2+b^2+c^2-2(ab+bc+ca)$ and $q^2$ being the momentum transfer squared bounded at $m_\ell^2\leq q^2\leq (m_{B^*}-m_P)^2$. For the four polarization vectors of  virtual $W^*$, $\bar{\epsilon}^{\mu}(\lambda_{W^*}=t,0,\pm)$, one can conveniently choose~\cite{Korner:1987kd,Korner:1989qb}
 \begin{eqnarray}\label{eq:polW}
\bar{\epsilon}^{\mu}(t)=\frac{1}{\sqrt{q^2}}(q_0,0,0,-|\vec{p}|)\,,\quad \bar{\epsilon}^{\mu}(0)=\frac{1}{\sqrt{q^2}}(|\vec{p}|,0,0,-q_0)\,,\quad  \bar{\epsilon}^{\mu}(\pm)=\frac{1}{\sqrt{2}}(0,\pm1,-i,0)\,.
\end{eqnarray}
Meanwhile, the polarization vectors of  initial $B^*$-meson could be written as
 \begin{eqnarray}\label{eq:polB}
\epsilon^{\mu}(0)=(0,0,0,1)\,,\quad  \epsilon^{\mu}(\pm)=\frac{1}{\sqrt{2}}(0,\mp1,-i,0)\,.
\end{eqnarray}

Turning to the $\ell-\bar{\nu}_\ell$ center-of-mass frame, the four momenta of lepton and antineutrino are given as
 \begin{eqnarray}
 p_\ell^{\mu}=(E_{\ell}, |\vec{p}_{\ell}|\sin\theta,0,|\vec{p}_{\ell}|\cos\theta)\,,\quad  p_{\nu_\ell}^{\mu}=(|\vec{p}_{\ell}|, -|\vec{p}_{\ell}|\sin\theta,0,-|\vec{p}_{\ell}|\cos\theta)
 \end{eqnarray}
where $E_{\ell}$ and $|\vec{p}_{\ell}|$ are the energy and the magnitude of the three-momentum of the charged lepton, respectively, given by $E_{\ell}=(q^2+m_{\ell}^2)/2\sqrt{q^2}$ and $|\vec{p}_{\ell}|=(q^2-m_{\ell}^2)/2\sqrt{q^2}$; and $\theta$ is the angle between the $P$ and ${\ell}$ three-momenta.  In this reference frame, the polarization vectors of  virtual $W^*$ take the form
 \begin{eqnarray}
\bar{\epsilon}^{\mu}(t)=(1,0,0,0)\,,\quad \bar{\epsilon}^{\mu}(0)=(0,0,0,1)\,, \quad \bar{\epsilon}^{\mu}(\pm)=\frac{1}{\sqrt{2}}(0,\mp1,-i,0)\,.
\end{eqnarray}

\subsection{Hadronic Helicity Amplitudes $H_{\lambda_{B^*}\lambda_{W^*}}$}
For  $\bar{B}^* \to P \ell^- \bar{\nu}_\ell$ decay, the hadronic helicity amplitude $H_{\lambda_{B^*} \lambda_{W^*}}$ defined by
 \begin{eqnarray}\label{eq:hha}
 H_{\lambda_{B^*} \lambda_{W^*}}=H_{\mu}(\lambda_{B^*})\,\bar{\epsilon}^{ *\mu}(\lambda_{W^*})
\end{eqnarray}
describes the decay of three helicity states of $B^*$ meson into a pseudo-scalar $P$ meson and the four helicity states of  virtual $W^*$. In Eq.~\eqref{eq:hha}, $H_{\mu}(\lambda_{B^*})$ represents hadronic matrix elements of the vector and axial-vector currents within the SM. For $B^*\to P$ transition, they are described by four QCD form factors $V(q^2)$ and $A_{0,1,2}(q^2)$ through
\begin{eqnarray}
\langle P(p_{P})|\bar{p}\gamma_{\mu} b|\bar{B}^*(\epsilon, p_{B^*})\rangle &=&
-\frac{2iV(q^2)}{m_{B^*}+m_{P}}\varepsilon_{\mu\nu\rho\sigma}
\epsilon^{\nu}p_{P}^{\rho}p_{B^*}^{\sigma},\\
\langle P(p_{P})|\bar{p}\gamma_{\mu}\gamma_5 b|\bar{B}^*(\epsilon, p_{B^*})\rangle &=&
2m_{B^*}A_0(q^2)\frac{\epsilon \cdot q}{q^2}q_{\mu}
+(m_{P}+m_{B^*})A_1(q^2)\left(\epsilon_{\mu}-\frac{\epsilon\cdot q}{q^2}q_{\mu}\right)\nonumber \\
&&+A_2(q^2)\frac{\epsilon \cdot q}{m_{P}+m_{B^*}} \left[(p_{B^*}+p_{P})_{\mu}-\frac{m^2_{B^*}-m^2_{P}} {q^2} q_{\mu}\right],
\end{eqnarray}
with the sign convention $\epsilon_{0123}=-1$.

Then, by contracting above hadronic matrix elements with the $B^*$ and $W^*$ polarization vectors given by Eqs.~\eqref{eq:polW} and \eqref{eq:polB}, we obtain four non-vanishing helicity amplitudes
\begin{eqnarray}
H_{0t}(q^2)&=&\frac{2m_{B^*}|\vec{p}|}{\sqrt{q^2}}A_0(q^2),\\
H_{00}(q^2)&=&\frac{1}{2m_{B^*}\sqrt{q^2}}\left[(m_{B^*}+m_{P})(m^2_{B^*}-m^2_{P}+q^2)A_1(q^2)
+ \frac{4m^2_{B^*}|\vec{p}|^2}{m_{B^*}+m_{P}}A_2(q^2) \right],\\
H_{\pm\mp}(q^2)&=&-(m_{B^*}+m_{P})A_1(q^2)\mp\frac{2m_{B^*}|\vec{p}|}{m_{B^*}+m_{P}}V(q^2).
\end{eqnarray}
It is obvious that only the amplitudes with $\lambda_{B^*}=\lambda_{P}-\lambda_{W^*}=-\lambda_{W^*}$ survive~\footnote{ Here, $\lambda_{W^*}=t$ has to be understood as $\lambda_{W^*}=0$ with $J=0$. }.

\subsection{ Helicity Amplitudes and Observables of $\bar{B}^* \to P \ell^- \bar{\nu}_\ell$ Decays}
Following the strategy of Refs.~\cite{Fajfer:2012vx,Korner:1987kd,Kadeer:2005aq}, one can expand the leptonic tensor in terms of a complete set of Wigner's $d^J$-functions.  As a result, $L_{\mu\nu}H^{\mu\nu}$ is reduced to a very compact form
\begin{eqnarray}\label{eq:ampd}
L_{\mu\nu}H^{\mu\nu}&=&\frac{1}{8} \sum_{\lambda_{\ell},\lambda_{\bar{\nu}_{\ell}}, \lambda_{W^*},\lambda_{W^*}^{\prime},\, J,\,J^{\prime}}\,(-1)^{J+J^{\prime}}\,|h_{\lambda_{\ell},\lambda_{\bar{\nu}_{\ell}}}|^2\,\delta_{\lambda_{B^*}\,,-\lambda_{W^*}}\,\delta_{\lambda_{B^*}\,,-\lambda_{W^*}^{\prime}}\nonumber\\
&&\times\, d^{J}_{\lambda_{W^*},\lambda_{\ell}-\frac{1}{2}}\,d^{J^{\prime}}_{\lambda_{W^*}^{\prime},\lambda_{\ell}-\frac{1}{2}}\,H_{\lambda_{B^*}\lambda_{W^*}}\,H_{\lambda_{B^*}\lambda_{W^*}^{\prime}}\,,
\end{eqnarray}
where $J$ and $J^{\prime}$ run over $1$ and $0$, $\lambda_{W^*}^{(\prime)}$ and $\lambda_{\ell}$ run over their components, and $\lambda_{\bar{\nu}_{\ell}}=\frac{1}{2}$. One may note that the non-diagonal interference contribution appears between the states of $J=1$, $\lambda_{W^*}=0$ and $J=0$, $\lambda_{W^*}=t$, but it has no contributions to the differential decay rate $d^2\Gamma/dq^2$ after integrating over $\cos \theta$, which can be seen from the following Eq.~\eqref{eq:DdGmp}.

The $h_{\lambda_{\ell},\lambda_{\bar{\nu}_{\ell}}}$ in Eq.~\eqref{eq:ampd} are the leptonic helicity amplitudes in the  $\ell-\bar{\nu}_\ell$ center-of-mass frame, and given by
\begin{eqnarray}
h_{\lambda_{\ell},\lambda_{\bar{\nu}_{\ell}}}=\bar{u}_{\ell}(\lambda_{\ell})\gamma^{\mu}(1-\gamma_5)\nu_{\bar{\nu}}(\frac{1}{2})\bar{\epsilon}_{\mu}(\lambda_{W^*})\,,
\end{eqnarray}
where $\lambda_{W^*}=\lambda_{\ell}-\lambda_{\bar{\nu}_{\ell}}$. The cases $\lambda_{\ell}=-1/2$ and $1/2$ are referred to as the non-flip and flip transitions, respectively. Taking the exact forms of the spinors and polarization vectors, we finally obtain  two nonvanishing contributions
\begin{eqnarray}
|h_{-\frac{1}{2},\frac{1}{2}}|^2&=&8(q^2-m_\ell^2)\,\quad \text{ non-flip}\,,\\
|h_{\frac{1}{2},\frac{1}{2}}|^2&=&8\frac{m_\ell^2}{2q^2}(q^2-m_\ell^2)\,\quad \text{ flip}\,,
\end{eqnarray}
which have exactly the same expressions as the one gotten in semileptonic $B$ and hyperon decays~\cite{Kadeer:2005aq,Fajfer:2012vx}.

By now, the basic building blocks of amplitudes have been obtained. Then, we present the observables considered in our following evaluations. The double differential decay rate of $\bar{B}^* \to P \ell^- \bar{\nu}_\ell$ decay could be written as
\begin{eqnarray}
\frac{d\Gamma}{dq^2d\cos\theta}=\frac{G_F^2|V_{pb}|^2}{(2\pi)^3}\,\frac{|\vec{p}|}{8m_{B^*}^2}\,\frac{1}{3}(1-\frac{m_\ell^2}{q^2})L_{\mu\nu}H^{\mu\nu}\,,
\end{eqnarray}
where the factor $1/3$ is caused by averaging over the spin of initial state $\bar{B}^*$. Further, using the standard convention for $d^J$-function~\cite{PDG14}, we finally obtain the double differential decay rates with a  given helicity state $(\lambda_{\ell}=\pm\frac{1}{2})$, which are
\begin{eqnarray}\label{eq:DdGml}
\frac{d^2\Gamma[\lambda_\ell=-1/2]}{dq^2d\cos\theta}&=&\frac{G_F^2|V_{pb}|^2|\vec{p}|}{256\pi^3m_{B^*}^2}\,\frac{1}{3}\,q^2\,(1-\frac{m_\ell^2}{q^2})^2\,\nonumber\\
&&\times\left[(1-\cos\theta)^2H_{-+}^2+(1+\cos\theta)^2H_{+-}^2+2\sin^2\theta H_{00}^2\right]\,,\\
\label{eq:DdGmp}
\frac{d^2\Gamma[\lambda_\ell=1/2]}{dq^2d\cos\theta}&=&\frac{G_F^2|V_{pb}|^2|\vec{p}|}{256\pi^3m_{B^*}^2}\,\frac{1}{3}\,q^2\,(1-\frac{m_\ell^2}{q^2})^2\,\frac{m_\ell^2}{q^2}\nonumber\\
&&\times\left[\sin^2\theta(H_{-+}^2+H_{+-}^2)+2(H_{0t}-\cos\theta H_{00})^2\right]\,.
\end{eqnarray}
Using Eqs.~\eqref{eq:DdGml} and \eqref{eq:DdGmp}, one can get the explicit forms of various observables of $\bar{B}^* \to P \ell^- \bar{\nu}_\ell$ decays.

Performing the integration over $\cos\theta$ and summing over the lepton helicity, we obtain the singly differential decay rate
\begin{eqnarray}\label{eq:SdG}
\frac{d\Gamma}{dq^2}=\frac{G_F^2|V_{pb}|^2|\vec{p}|}{96\pi^3m_{B^*}^2}\,\frac{1}{3}\,q^2\,(1-\frac{m_\ell^2}{q^2})^2\,\times\left[(H_{-+}^2+H_{+-}^2+H_{00}^2)(1+\frac{m_\ell^2}{2\,q^2})+\frac{3m_\ell^2}{2q^2}H_{0t}^2\right]\,,
\end{eqnarray}
from which the integrated decay rates, the branching fractions and the ratios defined by $R^*_{P}(q^2)$ $\equiv$ $ \frac{d\Gamma(\bar{B}^*\to P\tau^-\bar{\nu}_\tau)/dq^2}{d\Gamma(\bar{B}^*\to P\ell^{\prime-}\bar{\nu}_{\ell^{\prime}})/dq^2}$ $(\ell^{\prime}=\mu\,,e)$ are easily to be obtained. In addition, picking out the $H_{00}^2$ and $H_{0t}^2$ terms in Eq.~\eqref{eq:SdG}, one also can get the singly differential longitudinal decay rate $d\Gamma^L/dq^2$, as well as $R^{*L}_{P}(q^2)$, which are sensitive to the NP contributions of a charged scalar~\cite{Celis:2012dk}.
Besides the decay rate, there are also two important observables, the lepton spin asymmetry and the forward-backward asymmetry, which are defined as
\begin{eqnarray}
\label{eq:APLamb}
A^P_{\lambda}(q^2)&=&\frac{d\Gamma[\lambda_\ell=-1/2]/dq^2-d\Gamma[\lambda_\ell=1/2]/dq^2}{d\Gamma[\lambda_\ell=-1/2]/dq^2+d\Gamma[\lambda_\ell=1/2]/dq^2}\,,\\
A^P_{\theta}(q^2)&=&\frac{\int_{-1}^0d\cos\theta\,( d^2\Gamma/dq^2d\cos\theta)-\int_{0}^1d\cos\theta\,( d^2\Gamma/dq^2d\cos\theta)}{ d^2\Gamma/dq^2}\,,
\end{eqnarray}
respectively. In Eq.~\eqref{eq:APLamb}, the polarized differential decay rates $d\Gamma[\lambda_\ell=\pm1/2]/dq^2$ are obtained after integration over $\cos\theta$ of doubly differential ones given by Eqs.~\eqref{eq:DdGml} and~\eqref{eq:DdGmp}. Explicitly, we obtain
\begin{eqnarray}
A^P_{\lambda}(q^2)=\frac{(H_{00}^2+H_{-+}^2+H_{+-}^2)(1-\frac{m_\ell^2}{2\,q^2})-\frac{3m_\ell^2}{2q^2}H_{0t}^2}{(H_{00}^2+H_{-+}^2+H_{+-}^2)(1+\frac{m_\ell^2}{2\,q^2})+\frac{3m_\ell^2}{2q^2}H_{0t}^2}\,.
\end{eqnarray}
For $A^P_{\theta}(q^2)$, again using Eqs.~\eqref{eq:DdGml} and~\eqref{eq:DdGmp} and summing over the lepton helicity, we arrive at the explicit expression
\begin{eqnarray}
A^P_{\theta}(q^2)=\frac{3}{4}\frac{H_{-+}^2-H_{+-}^2+2\frac{m_\ell^2}{q^2}H_{00}H_{0t}}{(H_{00}^2+H_{-+}^2+H_{+-}^2)(1+\frac{m_\ell^2}{2\,q^2})+\frac{3m_\ell^2}{2q^2}H_{0t}^2}\,.
\end{eqnarray}

The lepton spin asymmetry $A_\lambda$ is very sensitive to the NP corrections, and therefore, has been widely studied in $B\to D^* \ell \bar{\nu_{\ell}}$ decays within various NP scenarios. However, unfortunately,  the lepton polarization can not be measured directly in the high energy experiments due to the lack of effective technology and method. For the case of $\tau$ lepton, its polarization could be determined  in principle through analyzing the full angular distribution of $\tau$ subsequent decay, but it is not very easy. Moreover, such way is not suitable for the case of light leptons~($\mu$ and $e$). It is hoped that the theoretical researches on $A_\lambda$ could motivate the development of the experimental technology and approach.

\section{Numerical Results and Discussions}
\subsection{Input Parameters}
Before presenting our predictions for $\bar{B}^* \to P \ell^- \bar{\nu}_\ell$ decays, we would like to clarify the input parameters used in our numerical evaluations.  For the CKM matrix elements, we use the fitted results $|V_{cb}|=41.80^{+0.33}_{-0.68}$ and $|V_{ub}|=3.714^{+0.072}_{-0.060}$ given by CKMFitter Group~\cite{Charles:2004jd}. For the well-known Fermi coupling constant $G_F$ and the masses of mesons and leptons, we take the averaged values given by PDG~\cite{PDG14}.

In order to  evaluate the branching fractions, the total decay widths (or  lifetimes) $\Gamma_{tot}(B^*)$ are essential. Due to the facts that there is no available experimental and theoretical information for $\Gamma_{\rm{tot}}(B^*)$ at present and the electromagnetic processes $B^*\to B\gamma$ dominate the decays of $B^*$ mesons, we take the approximation $\Gamma_{\rm{tot}}(B^*)\simeq \Gamma(B^*\to B\gamma)$ in our evaluations of branching fraction. The theoretical predictions on $\Gamma(B^*\to B\gamma)$ have been given in many different theoretical models~\cite{Goity:2000dk,Ebert:2002xz,Zhu:1996qy,Aliev:1995wi,Colangelo:1993zq,Choi:2007se,Cheung:2014cka}. In this paper, we will take the most recent results~\cite{Choi:2007se,Cheung:2014cka}
\begin{eqnarray} 
\label{eq:GtotBu}
\Gamma_{\rm{tot}}(B^{*+})&\simeq& \Gamma(B^{*+}\to B^+ \gamma)=(468^{+73}_{-75})\,{\rm eV},\\
 \label{eq:GtotBd}
\Gamma_{\rm{tot}}(B^{*0})&\simeq& \Gamma(B^{*0}\to B^0 \gamma)=(148\pm20)\,{\rm eV},\\
 \label{eq:GtotBs}
\Gamma_{\rm{tot}}(B^{*0}_s)&\simeq& \Gamma(B^{*0}_s\to B^0_s \gamma)=(68\pm17)\,{\rm eV}.
\end{eqnarray}

\begin{table}[t]
\caption{The values of form factors $A_{0,1,2}(0)$ and $V(0)$ within BSW model.}
\begin{center}
\begin{tabular}{lccccc}
\hline\hline
Transition                  &$A_0(0)$       &$A_1(0)$    &$A_2(0)$    &$V(0)$ \\
\hline
$\bar{B}^{*} \to D$         &$0.63$         &$0.66$      &$0.56$      &$0.70$   \\
$\bar{B}^{*}_s \to D_s$  &$0.59$         &$0.61$      &$0.54$      &$0.67$   \\
\hline
$\bar{B}^{*} \to \pi$       &$0.34$         &$0.38$      &$0.29$      &$0.34$   \\
$\bar{B}^{*}_s \to K$    &$0.29$         &$0.31$      &$0.28$      &$0.32$   \\
\hline\hline
\end{tabular}
\end{center}
\label{tab:formfactor}
\end{table}

Besides, the transition form factors are also essential inputs, but no ready-made results could be used at present. In this paper, the Bauer-Stech-Wirbel~(BSW) model~\cite{Wirbel:1985ji,Bauer:1988fx} is employed for evaluating the  form factors.
Within the BSW framework,  the form factors $A_{0,1,2}(q^2)$ and $V(q^2)$ for $\bar{B}^*\to P$ transitions at $q^2=0$ could be written as the overlap integrals of wave functions of mesons~\cite{Wirbel:1985ji},
\begin{eqnarray}
A^{\bar{B}^*\to P}_0(0)&=&\int{d^2 p_{\perp}}\int_0^1{dx \varphi_{P}(\vec{p}_{\perp},x)\sigma_z \varphi^{1,0}_{\bar{B}^*}(\vec{p}_{\perp},x)},  \\
A^{\bar{B}^*\to P}_1(0)&=&\frac{m_b+m_c}{m_{\bar{B}^*}+m_{P}}J^{\bar{B}^*\to P},\\
A^{\bar{B}^*\to P}_2(0)&=&\frac{2m_{\bar{B}^*}}{m_{\bar{B}^*}-m_{P}}A^{\bar{B}^*\to P}_0(0)-\frac{m_{\bar{B}^*}+m_{P}}{m_{\bar{B}^*}-m_{P}}A^{\bar{B}^*\to P}_1(0),\\
V^{\bar{B}^*\to P}(0)&=&\frac{m_b-m_c}{m_{\bar{B}^*}-m_{P}}J^{\bar{B}^*\to P},\\
J^{\bar{B}^*\to P}&=&\sqrt{2}\int{d^2p_{\perp}}\int^1_0{dx \varphi_{P}(\vec{p}_{\perp},x)\sigma_y \varphi^{1,-1}_{\bar{B}^*}(\vec{p}_{\perp},x)}\,,
\end{eqnarray}
where $\vec{p}_{\perp}$ is the transverse quark momentum. With the meson wave function $\varphi(\vec{p}_{\perp},x)$ as solution of a relativistic scalar harmonic oscillator potential~\cite{Wirbel:1985ji}, using the constituent masses $m_u=m_d=0.39\,{\rm GeV}$, $m_s=0.50\,{\rm GeV}$, $m_c=1.62\,{\rm GeV}$, $m_b= 4.94\,{\rm GeV}$ and $\omega=\sqrt{\langle\vec{p}_{\perp}^2\rangle}=0.4\,{\rm GeV}$, we obtain the numerical results of the form factors at $q^2=0$, which are summarized in Table~\ref{tab:formfactor}. To be conservative,  in our following evaluation, we assign $15\%$ uncertainties to these values.
Moreover, with the assumption of the nearest pole dominance,  the dependences of  form factors on $q^2$  are explicitly written as~\cite{Wirbel:1985ji,Bauer:1988fx}
\begin{eqnarray}
A^{\bar{B}^*\to P}_0(q^2)&\simeq&\frac{A_0(0)}{1-q^2/m^2_{B_p(0^-)}}\,, \quad
A^{\bar{B}^*\to P}_1(q^2)\simeq\frac{A_1(0)}{1-q^2/m^2_{B_p(1^+)}}, \nonumber \\
A^{\bar{B}^*\to P}_2(q^2)&\simeq&\frac{A_2(0)}{1-q^2/m^2_{B_p(1^+)}}\,, \quad
V^{\bar{B}^*\to P}(q^2)\simeq\frac{V(0)}{1-q^2/m^2_{B_p(1^-)}},
\end{eqnarray}
where $B_p(J^P)$ is the state of $B_p$ with quantum number of $J^P$~($J$ and $P$ are the quantum numbers of total angular momenta and parity, respectively).
 In addition, it should be noted that, instead of using the BSW model, a particularly convenient parameterization of the form factors has been obtained by using dispersion relations in QCD and the heavy quark symmetry~\cite{Caprini:1997mu}, which is widely used for the study of $B\to D^{(*)}\ell\bar{\nu}_{\ell}$ decays. Especially, further combining with the Lattice QCD~(LQCD) calculation at high $q^2$, {\it e.g.} Ref.~\cite{deDivitiis:2008df}, one may get much more reliable results of hadronic form factors. So, for the  $B^{*}\to D\ell\bar{\nu}_{\ell}$ decays, once the LQCD results relevant to $B^{*}\to D$ transition are available in the future, much more accurate and reliable theoretical predictions for the observables are expected. 

\subsection{Theoretical Prediction and Discussion}

\begin{table}[t]
\caption{The theoretical predictions for the branching fractions of  $ B^*\to P \ell^- \bar{\nu}_\ell$ decays.}
\begin{center}
\begin{tabular}{lc|lc}
\hline\hline
Decay mode                                   &$\mathcal{B}$ & Decay mode                                   &$\mathcal{B}$\\\hline
$\bar{B}^{*-} \to \pi^0 \ell'^- \bar{\nu}_{\ell'}$
  &$2.02$$^{+0.67}_{-0.57}$$^{+0.08}_{-0.06}$$^{+0.37}_{-0.28}$$\times10^{-10}$
&  $\bar{B}^{*-} \to D^0 \ell'^- \bar{\nu}_{\ell'}$
  &$2.29$$^{+0.72}_{-0.61}$$^{+0.04}_{-0.07}$$^{+0.42}_{-0.32}$$\times10^{-8}$\\\hline
$\bar{B}^{*-} \to \pi^0 \tau^- \bar{\nu}_{\tau}$
  &$1.37$$^{+0.45}_{-0.39}$$^{+0.05}_{-0.04}$$^{+0.25}_{-0.19}$$\times10^{-10}$
 & $\bar{B}^{*-} \to D^0\tau^- \bar{\nu}_{\tau}$
  &$6.83$$^{+2.06}_{-1.75}$$^{+0.11}_{-0.22}$$^{+1.26}_{-0.94}$$\times10^{-9}$\\\hline
$\bar{B}^{*0} \to \pi^+ \ell'^- \bar{\nu}_{\ell'}$
  &$1.28$$^{+0.43}_{-0.36}$$^{+0.05}_{-0.04}$$^{+0.20}_{-0.15}$$\times10^{-9}$
  &$\bar{B}^{*0} \to D^+ \ell'^- \bar{\nu}_{\ell'}$
  &$7.20$$^{+2.23}_{-1.95}$$^{+0.11}_{-0.23}$$^{+1.13}_{-0.86}$$\times10^{-8}$\\\hline
$\bar{B}^{*0} \to \pi^+ \tau^- \bar{\nu}_{\tau}$
  &$8.63$$^{+2.86}_{-2.42}$$^{+0.34}_{-0.28}$$^{+1.35}_{-1.02}$$\times10^{-10}$
  &$\bar{B}^{*0} \to D^+ \tau^- \bar{\nu}_{\tau}$
  &$2.14$$^{+0.64}_{-0.57}$$^{+0.03}_{-0.07}$$^{+0.33}_{-0.25}$$\times10^{-8}$\\\hline
$\bar{B}^{*0}_s \to K^+ \ell'^- \bar{\nu}_{\ell'}$
  &$1.67$$^{+0.57}_{-0.49}$$^{+0.07}_{-0.05}$$^{+0.56}_{-0.33}$$\times10^{-9}$
  &$\bar{B}^{*0}_s \to D^+_s \ell'^- \bar{\nu}_{\ell'}$
  &$1.39$$^{+0.43}_{-0.37}$$^{+0.02}_{-0.04}$$^{+0.46}_{-0.28}$$\times10^{-7}$\\\hline
$\bar{B}^{*0}_s \to K^+ \tau^- \bar{\nu}_{\tau}$
  &$1.07$$^{+0.35}_{-0.31}$$^{+0.04}_{-0.03}$$^{+0.36}_{-0.21}$$\times10^{-9}$
  &$\bar{B}^{*0}_s \to D^+_s \tau^- \bar{\nu}_{\tau}$
  &$4.08$$^{+1.24}_{-1.08}$$^{+0.06}_{-0.13}$$^{+1.36}_{-0.82}$$\times10^{-8}$\\\hline
\hline
\end{tabular}
\end{center}
\label{tab:Bstar2P}
\end{table}

\begin{table}[t]
\caption{Predictions for $q^2$-integrated observables  $A_{\lambda\,,\theta}^P$~($\ell=\tau$) and $R^{*(L)}_P$.}
\begin{center}
\begin{tabular}{lc|lc|lc}
\hline\hline
Obs.                      &Prediction                      &Obs.                   &Prediction                      &Obs.                          &Prediction   \\\hline                            
$A^D_{\lambda}$     &$0.546^{+0.043}_{-0.069}$     &$A^{\pi}_{\lambda}$   &$0.767^{+0.024}_{-0.008}$ &$A^K_{\lambda}$  &$0.734^{+0.031}_{-0.010}$\\\hline
$A^D_{\theta}$      &$0.106^{+0.031}_{-0.026}$        &$A^{\pi}_{\theta}$     &$0.049^{+0.001}_{-0.010}$  &$A^K_{\theta}$       &$0.060^{+0.002}_{-0.012}$\\\hline
$R^*_D$                 &$0.298^{+0.001}_{-0.001}$        &$R^*_{\pi}$                &$0.677^{+0.016}_{-0.016}$   &$R^*_K$                 &$0.637^{+0.019}_{-0.018}$\\\hline
$R^{*L}_D$            &$0.254^{+0.009}_{-0.004}$       &$R^{*L}_{\pi}$           &$0.651^{+0.040}_{-0.025}$ &$R^{*L}_K$              &$0.609^{+0.047}_{-0.028}$\\\hline
\hline
\end{tabular}
\end{center}
\label{tab:Obs}
\end{table}

\begin{figure}[t]
\caption{The $q^2$ dependence of differential decay rates $d\Gamma/dq^2$ (solid lines) and $d\Gamma^L/dq^2$ (dashed lines).}
\begin{center}
\subfigure[]{\includegraphics[scale=0.6]{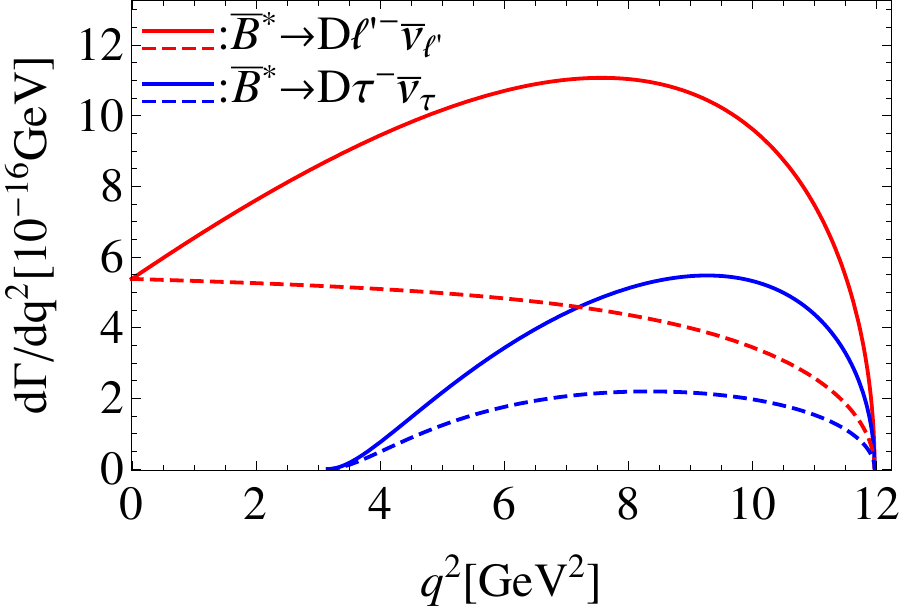}}\qquad\quad
\subfigure[]{\includegraphics[scale=0.6]{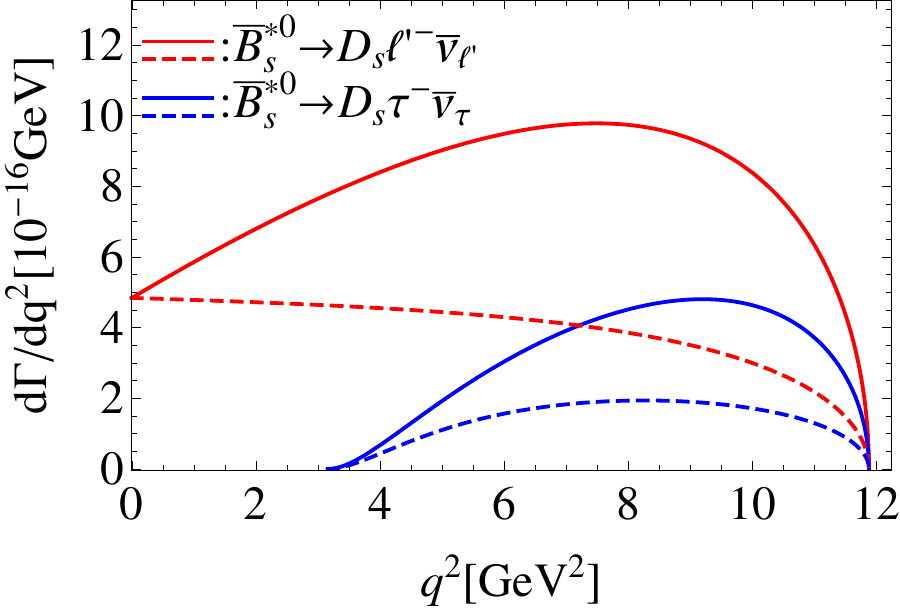}}\\
\subfigure[]{\includegraphics[scale=0.6]{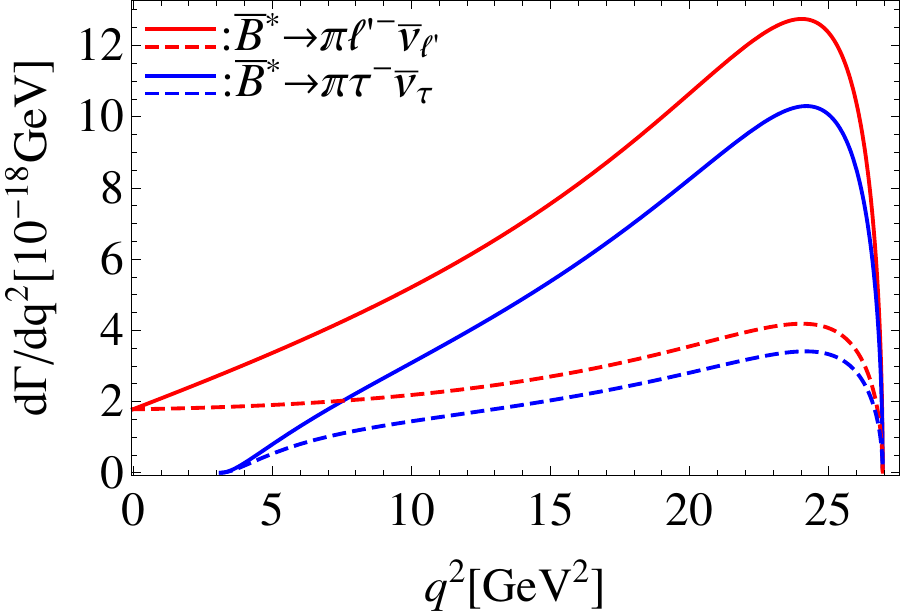}}\qquad\quad
\subfigure[]{\includegraphics[scale=0.6]{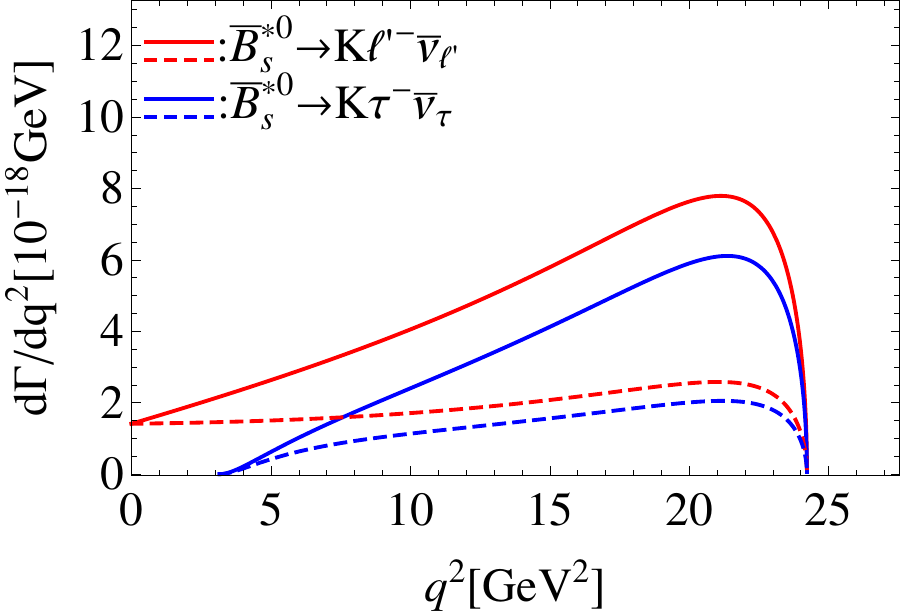}}
\end{center}
\label{fig:dG}
\end{figure}
\begin{figure}[t]
\caption{The $q^2$-dependence of the observables $R^{*(L)}_P$ and $A_{\lambda\,,\theta}^P$ of $ \bar{B}^* \to P \ell^- \bar{\nu}_\ell$  decays .}
\begin{center}
\subfigure[]{\includegraphics[scale=0.6]{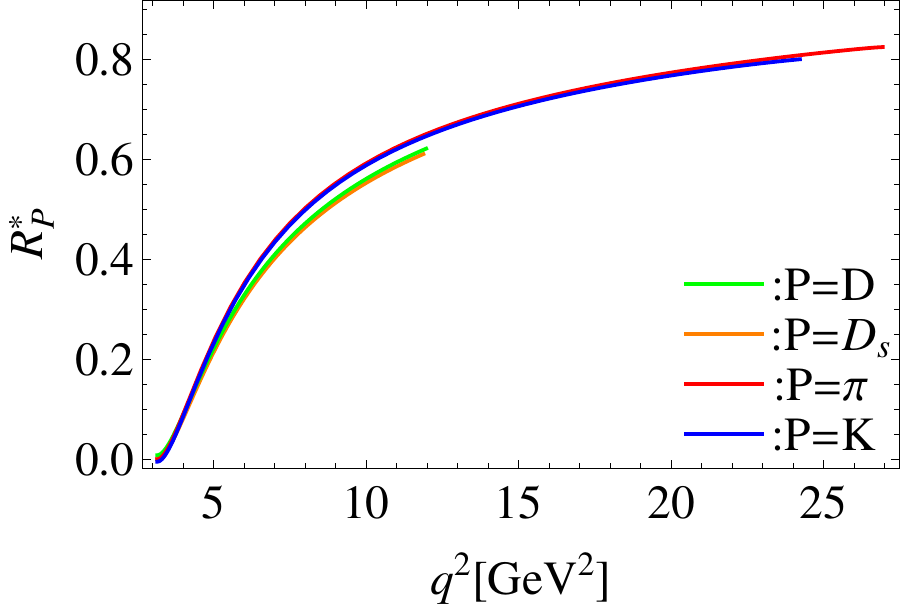}}\qquad\quad
\subfigure[]{\includegraphics[scale=0.6]{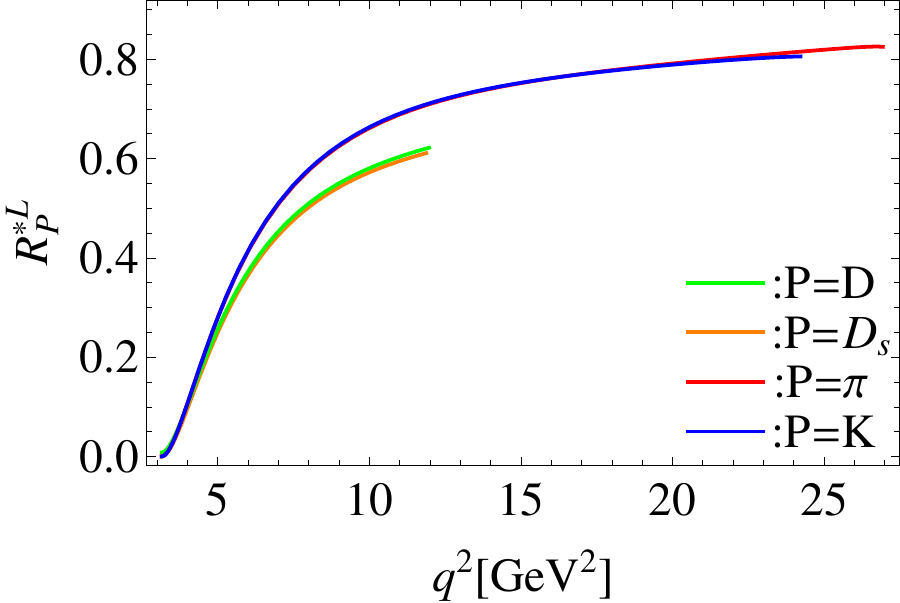}}\\
\subfigure[]{\includegraphics[scale=0.6]{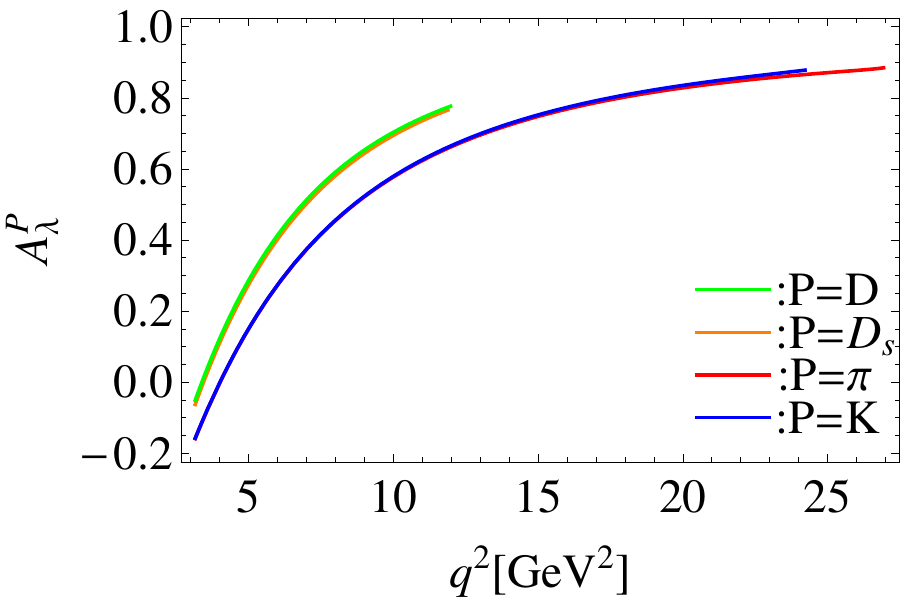}}\qquad\quad
\subfigure[]{\includegraphics[scale=0.6]{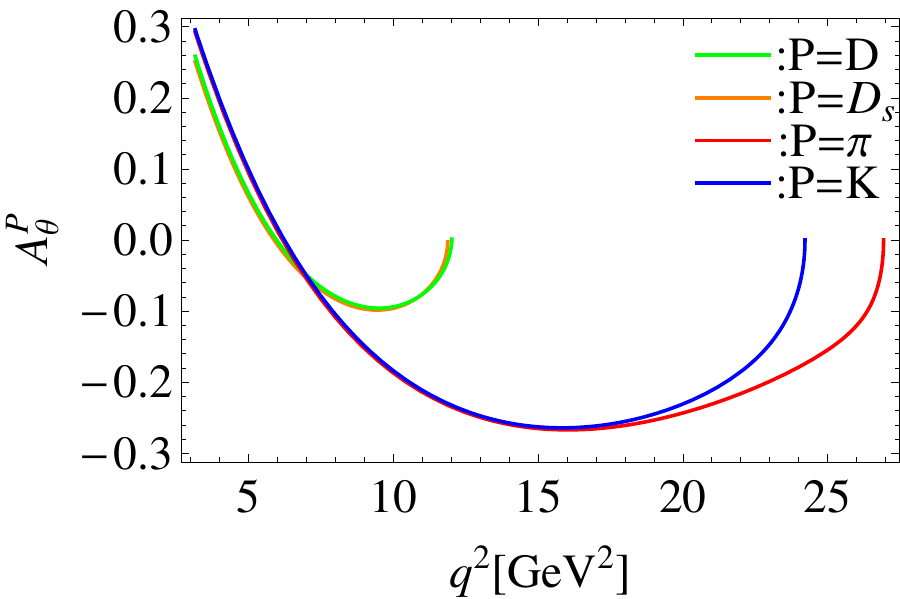}}
\end{center}
\label{fig:dobs}
\end{figure}

With the input values and the formula given above, we then present our theoretical predictions and discussion. In Table~\ref{tab:Bstar2P}, we summarize the predictions of branching fractions, in which the three theoretical errors are caused by the uncertainties of form factors, CKM factors and $\Gamma_{\rm tot}(B^*)$, respectively. For the other $q^2$-integrated observables $A_{\lambda\,,\theta}^P$~($\ell=\tau$) and $R^{*(L)}_P$, the predictions are given in Table~\ref{tab:Obs}, in which the theoretical uncertainties are caused by the  form factors only. In Figs. \ref{fig:dG} and \ref{fig:dobs}, the $q^2$-dependence of differential decay rates $d\Gamma^{(L)}/dq^2$ and  $A_{\lambda\,,\theta}^P$, $R^{*(L)}_P$ are shown, respectively. The following are some discussions:
\begin{enumerate}
\item[(1)] Compared with $\bar{B}^*_{(s)} \to D_{(s)} \ell^- \bar{\nu}_{\ell}$ decays, $\bar{B}^*_{(s)} \to \pi(K) \ell^- \bar{\nu}_{\ell}$ decays are suppressed by both an additional factor $\lambda$ and the relatively small form factors. Therefore, the branching fractions of $\bar{B}^*_{(s)} \to \pi(K) \ell^- \bar{\nu}_{\ell}$ decays are expected to be much smaller than the ones of corresponding $\bar{B}^*_{(s)} \to D_{(s)} \ell^- \bar{\nu}_{\ell}$ decays by a factor of ${\cal O}(10^{-1})\sim{\cal O}(10^{-2})$, which can be seen from Table~\ref{tab:Bstar2P}.

 In addition, from Table~\ref{tab:Bstar2P}, it could also be found that ${\cal B}(B^{*-} \to D^0 \ell^- \bar{\nu}_{\ell})$ $:$ ${\cal B}(\bar{B}^{*0} \to D^+ \ell^- \bar{\nu}_{\ell})$ $:$ ${\cal B}(\bar{B}^*_{s} \to D_{s} \ell^- \bar{\nu}_{\ell})$ $\approx$ $1$ $:$ $2$ $:$ $6$, which is mainly attributed to the  total decay widths $\Gamma_{tot}(B^*)$ illustrated by Eqs.~\eqref{eq:GtotBu}, \eqref{eq:GtotBd} and \eqref{eq:GtotBs}.

\item[(2)]
In Table~\ref{tab:Bstar2P}, one may find that ${\cal B}(\bar{B}^*_{(s)} \to \pi(K) \ell^- \bar{\nu}_{\ell}) \lesssim 10^{-9}$, which implies that $\bar{B}^*_{(s)} \to \pi(K) \ell^- \bar{\nu}_{\ell}$ decays are hardly to be observed by Belle-II. However, fortunately, all of $\bar{B}^* \to D \ell^- \bar{\nu}_{\ell}$ decay modes are in the scope of SuperKEKB/Belle-II experiment due to ${\cal B}(\bar{B}^* \to D \ell^- \bar{\nu}_{\ell}) > 10^{-9}$, in which $\bar{B}^*_{s} \to D_{s} \ell^{\prime -} \bar{\nu}_{\ell^{\prime}}$ decay has the largest branching fraction of the order $\sim$ ${\cal O}(10^{-7})$, and therefore, should be sought for with priority and firstly observed. 

 Moreover, $\bar{B}^* \to D \ell^- \bar{\nu}_{\ell}$ decay modes are also expected to be measured by LHC experiments, which can be seen from the following rough analysis. Here, we take the possible measurement of $\bar{B}^{*0} \to D^+ \ell'^- \bar{\nu}_{\ell'}$ decay at LHCb as an example. Firstly, it is expected that about $2\times 50/3\times3.63\times10^{5}=1.21\times 10^{7}$ $\bar{B}^0 \to D^{*+} \mu^- \bar{\nu}_{\mu}$ decay events  will be found after LHCb   upgrade due to the facts that (i) using the data corresponding to integrated luminosities of $1.0\,{\rm fb}^{-1}$ and $2.0\,{\rm fb}^{-1}$ collected at $pp$ center-of-mass energy $\sqrt{s}=7$ and $8\,{\rm TeV}$, respectively, $3.63\times10^{5}$ $\bar{B} \to D^{*+} \mu^- \bar{\nu}_{\mu}$ decay events have been found by LHCb collaboration~\cite{Aaij:2015yra};  (ii) After high-luminosity upgrade, a data sample of $50\,{\rm fb}^{-1}$ will be collected  by LHCb collaboration at a much higher $\sqrt{s}=14\,{\rm TeV}$, which will results in a further enhancement of $b\bar{b}$ production by a factor about 2~\cite{Bediaga:2012py,LHCb:upgrade}. Secondly, one can assume that the most of $B$ mesons detected at LHC are mainly produced through $B^*\to B\gamma$ decay because $B^*$ mesons are often produced by about 3 times more than the $B$ mesons, which has been confirmed by the measurements at $Z^0$ peak by LEP~\cite{Buskulic:1995mt}. Finally,  further taking ${\cal B}(\bar{B}^{*0} \to D^+ \ell'^- \bar{\nu}_{\ell'})/{\cal B}(\bar{B}^0 \to D^{*+} \mu^- \bar{\nu}_{\mu}) \sim 1.4\times 10^{-6}$ into account, one may estimate that about ${\cal O}(10)$ $\bar{B}^{*0} \to D^+ \ell'^- \bar{\nu}_{\ell'}$ events could be observed by LHCb. In addition, if one take $\bar{B}^0_s\to \mu\mu$  instead of $\bar{B}^0 \to D^{*+} \mu^- \bar{\nu}_{\mu}$  as a reference and revisit the estimation above,  it can be found that about ${\cal O}(10^4)$ $\bar{B}^{*0} \to D^+ \ell'^- \bar{\nu}_{\ell'}$ decay events are expected to be observed in the high-luminosity LHC era. 

 \item[(3)]
Recalling the $\bar{B} \to D^{(*)} \ell \bar{\nu}_{\ell}$ decays, the known ``$R_{D^{(*)}}$ puzzles''  provide possible hints towards NP, especially the one of lepton flavor (universality) violation. If it is the truth,  the corresponding NP corrections should also affect $\bar{B}^{*} \to D \ell \bar{\nu}_{\ell}$ decays, and therefore, the future measurements for $R_{D}^{*(L)}$ should significantly deviate from the SM results. Otherwise, the NP models providing solutions to  
  ``$R_{D^{(*)}}$ puzzles'' will suffer a serious challenge from $R_{D}^{*(L)}$. So, the future measurements of $R_{D}^{*(L)}$ will play an important role for testing the SM and the various NP models.
  
To distinguish the possible NP hints, it will become important to control the theoretical uncertainties as well as possible.
From our predictions for  $R_{D}^{*(L)}$ given in Table~\ref{tab:Obs}, as expected, one may find that the uncertainty caused by the hadronic factors is significantly reduced compared to the decay rates. Moreover, when the range of $q^2$ integration is the same in the numerator and the denominator of $R_{D}^{*}$, the cancellation of the nonperturbative error further improves, allowing for more precise predictions of the ratio of partial rates~\cite{Bernlochner:2015mya,Tanaka:1994ay}. Numerically, for instance, choosing the $q^2$ integration range $[m^2_{\tau},q^2_{\rm max}]$ for both  numerator and  denominator,  we get
\begin{eqnarray*}
\widetilde{R}_{D}^{*}\equiv\frac{\int^{q^2_{\rm max}}_{m^2_{\tau}}{\rm d}q^2 d \Gamma(\bar{B}^* \to D \tau^- \nu_{\tau})/{\rm d}q^2}{\int^{q^2_{\rm max}}_{m^2_{\tau}}{\rm d}q^2 d \Gamma(\bar{B}^* \to D \ell^{\prime -} \nu_{\ell^{\prime}})/{\rm d}q^2}=0.378\,,
 \end{eqnarray*}
which could be measured with a lower cut on $q^2$. 
In addition,  the $q^2$-dependences of $R_{D}^{*(L)}$ are shown in Figs.~\ref{fig:dobs} (a) and (b), which, once measured,  would present a much stricter test for the SM and  NP. 
 
 \item[(4)]
 For the lepton spin asymmetry and the forward-backward asymmetry, our numerical results are listed in Table~\ref{tab:Obs}. Similar to $R_{D}^{*(L)}$, because of the cancellation of the hadronic errors between numerator and denominator,  the theoretical uncertainties are significantly small compared with the branching fraction. 
 Regarding their differential distributions, which are shown in Figs.~\ref{fig:dobs} (c) and (d), a characteristic feature is the zero-crossing point, which is usually used to distinguish the NP effects from the SM, or different NP scenarios. Numerically, we get that $A^P_{\lambda}(q^2)$ and $A^P_{\theta}(q^2)$  cross the zero point respectively at $q^2=3.4\,{\rm GeV}$ and $5.8\,{\rm GeV}$ for $P=D$, and $q^2=4.0\,{\rm GeV}$ and $6.2\,{\rm GeV}$ for $P=\pi\,,K$.
\end{enumerate}

\section{Summary}
The $B^*$ weak decays are legal within the Standard Model, although their branching ratios are tiny compared with the electromagnetic decays. In  this paper, motivated by abundant $B^*$ data samples at high-luminosity heavy-flavor experiments in the future, we have studied the tree-dominated semileptonic $\bar{B}^*_{u,d,s} \to P \ell^- \bar{\nu}_\ell$ ($P=D\,,D_s\,,\pi\,,K$ and $\ell=\tau\,,\mu\,,e$) decays within the Standard Model. The helicity amplitudes are calculated in detail, and the predictions of observables including branching fraction~(decay rate),  lepton spin asymmetry, forward-backward asymmetry and ratio $R_D^{\ast(L)}$ are firstly presented in Tables~\ref{tab:Bstar2P}, \ref{tab:Obs} and  Figs. \ref{fig:dG}, \ref{fig:dobs}. It is found that the CKM-favored $\bar{B}^* \to D \ell^- \bar{\nu}_\ell$ decays have relatively large branching fractions of ${\cal O}(10^{-9})$$\sim$${\cal O}(10^{-7})$, and hence are hopefully to be measured by the heavy-flavor experiments at running LHC and forthcoming SuperKEKB/Belle-II. 

\section*{Acknowledgments}
  We thank Yue-Hong Xie, Ya-Dong Yang and Xin-Qiang Li at CCNU, Hai-Bo Li at IHEP and Nan Li at HNNU for helpful discussion and comments.  The work is supported by the National Natural  Science Foundation of China (Grant Nos. 11547014, 11475055 and 11275057).
  Q. Chang is also supported by the Foundation  for the Author of National Excellent Doctoral  Dissertation of P. R. China (Grant No. 201317),  the Program for Science and Technology Innovation  Talents in Universities of Henan Province  (Grant No. 14HASTIT036).



\begin{thebibliography}{99}
 \bibitem{HFAG}
  Y.~Amhis {\it et al.} [Heavy Flavor Averaging Group (HFAG) Collaboration],
  arXiv:1412.7515 [hep-ex],  online update at: http://www.slac.stanford.edu/xorg/hfag.

   \bibitem{Charles:2004jd}
 J. Charles {\it et al.} (CKMfitter Group), Eur. Phys. J. C {\bf 41} (2005) 1;
 updated results and plots available at: http://ckmfitter.in2p3.fr.

   \bibitem{Lees:2012xj}
  J.~P.~Lees {\it et al.} [BaBar Collaboration],
  Phys.\ Rev.\ Lett.\  {\bf 109} (2012) 101802.

  \bibitem{Lees:2013uzd}
  J.~P.~Lees {\it et al.} [BaBar Collaboration],
  Phys.\ Rev.\ D {\bf 88} (2013) no. 7, 072012.

  \bibitem{Huschle:2015rga}
  M.~Huschle {\it et al.} [Belle Collaboration],
  Phys.\ Rev.\ D {\bf 92} (2015) no.7,  072014.

  \bibitem{Kuhr:2015muu}
  T.~Kuhr [Belle Collaboration],
  PoS FPCP {\bf 2015} (2015) 015.

  \bibitem{Abdesselam:2016cgx}
  A.~Abdesselam {\it et al.} [The Belle Collaboration],
  arXiv:1603.06711 [hep-ex].

\bibitem{Aaij:2015yra}
  R.~Aaij {\it et al.} [LHCb Collaboration],
  Phys.\ Rev.\ Lett.\  {\bf 115} (2015) no.11,  111803
   Addendum: [Phys.\ Rev.\ Lett.\  {\bf 115} (2015) no.15,  159901].

 \bibitem{Fajfer:2012vx}
  S.~Fajfer, J.~F.~Kamenik and I.~Nisandzic,
  Phys.\ Rev.\ D {\bf 85} (2012) 094025.

  \bibitem{Lattice:2015rga}
  J.~A.~Bailey {\it et al.} [MILC Collaboration],
  Phys.\ Rev.\ D {\bf 92} (2015) no.3,  034506.

  \bibitem{Na:2015kha}
  H.~Na {\it et al.} [HPQCD Collaboration],
  Phys.\ Rev.\ D {\bf 92} (2015) no.5,  054510.

\bibitem{Fan:2015kna}
  Y.~Y.~Fan, Z.~J.~Xiao, R.~M.~Wang and B.~Z.~Li,
  Science Bulletin Vol. 60 (2015) 2009-2015.

\bibitem{Fan:2014}
  Y. Y. Fan, W. F. Wang, Shan Cheng and Z. J. Xiao, Science Bulletin Vol. 59 (2014) 125-132.

  \bibitem{Fajfer:2012jt}
  S.~Fajfer, J.~F.~Kamenik, I.~Nisandzic and J.~Zupan,
  Phys.\ Rev.\ Lett.\  {\bf 109} (2012) 161801.

\bibitem{Sakaki:2012ft}
  Y.~Sakaki and H.~Tanaka,
  Phys.\ Rev.\ D {\bf 87} (2013) no.5,  054002.

\bibitem{Datta:2012qk}
  A.~Datta, M.~Duraisamy and D.~Ghosh,
  Phys.\ Rev.\ D {\bf 86} (2012) 034027.

\bibitem{Bailey:2012jg}
  J.~A.~Bailey {\it et al.},
  Phys.\ Rev.\ Lett.\  {\bf 109} (2012) 071802.

\bibitem{Becirevic:2012jf}
  D.~Becirevic, N.~Kosnik and A.~Tayduganov,
  Phys.\ Lett.\ B {\bf 716} (2012) 208.

\bibitem{Tanaka:2012nw}
  M.~Tanaka and R.~Watanabe,
  Phys.\ Rev.\ D {\bf 87} (2013) no.3,  034028.

\bibitem{Freytsis:2015qca}
  M.~Freytsis, Z.~Ligeti and J.~T.~Ruderman,
  Phys.\ Rev.\ D {\bf 92} (2015) no.5,  054018

\bibitem{Bhattacharya:2015ida}
  S.~Bhattacharya, S.~Nandi and S.~K.~Patra,
  Phys.\ Rev.\ D {\bf 93} (2016) no.3,  034011.

\bibitem{Celis:2012dk}
  A.~Celis, M.~Jung, X.~Q.~Li and A.~Pich,
  JHEP {\bf 1301} (2013) 054.

\bibitem{Ko:2012sv}
  P.~Ko, Y.~Omura and C.~Yu,
  JHEP {\bf 1303} (2013) 151.

\bibitem{Crivellin:2012ye}
  A.~Crivellin, C.~Greub and A.~Kokulu,
    Phys.\ Rev.\ D {\bf 86} (2012) 054014.

\bibitem{Deshpande:2012rr}
  N.~G.~Deshpande and A.~Menon,
  JHEP {\bf 1301} (2013) 025.

\bibitem{Sakaki:2013bfa}
  Y.~Sakaki, M.~Tanaka, A.~Tayduganov and R.~Watanabe,
  Phys.\ Rev.\ D {\bf 88} (2013) no.9,  094012.

  \bibitem{Greljo:2015mma}
  A.~Greljo, G.~Isidori and D.~Marzocca,
  JHEP {\bf 1507} (2015) 142.

\bibitem{Dorsner:2013tla}
  I.~Dorsner, S.~Fajfer, N.~Kosnik and I.~Nisandzic,
  JHEP {\bf 1311} (2013) 084.

\bibitem{Biancofiore:2013ki}
  P.~Biancofiore, P.~Colangelo and F.~De Fazio,
  Phys.\ Rev.\ D {\bf 87} (2013) no.7,  074010.

  \bibitem{Bauer:2015knc}
  M.~Bauer and M.~Neubert,
  arXiv:1511.01900 [hep-ph].

  \bibitem{Fajfer:2015ycq}
  S.~Fajfer and N.~Kosnik,
  Phys.\ Lett.\ B {\bf 755} (2016) 270.

  \bibitem{Hati:2016thk}
  C.~Hati,
  arXiv:1601.02457 [hep-ph].

  \bibitem{Zhu:2016xdg}
  J.~Zhu, H.~M.~Gan, R.~M.~Wang, Y.~Y.~Fan, Q.~Chang and Y.~G.~Xu,
  arXiv:1602.06491 [hep-ph].

  \bibitem{Alonso:2016gym}
  R.~Alonso, A.~Kobach and J.~M.~Camalich,
  arXiv:1602.07671 [hep-ph].

  \bibitem{Isgur:1991wq}
  N.~Isgur and M.~B.~Wise,
  Phys.\ Rev.\ Lett.\  {\bf 66} (1991) 1130.

\bibitem{Godfrey:1986wj}
  S.~Godfrey and R.~Kokoski,
  Phys.\ Rev.\ D {\bf 43} (1991) 1679.

\bibitem{Eichten:1993ub}
  E.~J.~Eichten, C.~T.~Hill and C.~Quigg,
  Phys.\ Rev.\ Lett.\  {\bf 71} (1993) 4116.

\bibitem{Ebert:1997nk}
  D.~Ebert, V.~O.~Galkin and R.~N.~Faustov,
  Phys.\ Rev.\ D {\bf 57} (1998) 5663
   [Erratum Phys.\ Rev.\ D {\bf 59} (1998) 019902].

  \bibitem{Abe:2010gxa}
  T.~Abe {\it et al.}  [Belle-II Collaboration], arXiv:1011.0352.

  \bibitem{Huang:2006mf}
  G.~S.~Huang {\it et al.}  [CLEO Collaboration], hep-ex/0607080.

  \bibitem{PDG14}
  K.~A.~Olive {\it et al.} [Particle Data Group Collaboration],
  Chin.\ Phys.\ C {\bf 38}  (2014) 090001.

    \bibitem{Grinstein:2015aua}
  B.~Grinstein and J.~M.~Camalich,
  arXiv:1509.05049 [hep-ph].

    \bibitem{Aaij:2010gn}
  R. Aaij {\it et al.}  (LHCb Collaboration),  Phys.  Lett. B {\bf 694} (2010) 209.

 \bibitem{Bediaga:2012py}
  R.~Aaij {\it et al.} [LHCb Collaboration],
  Eur.\ Phys.\ J.\ C {\bf 73} (2013) no. 4, 2373.

    \bibitem{Aaij:2014jba}
  R. Aaij {\it et al.}  (LHCb Collaboration), Int. J. Mod. Phys. A {\bf 30} (2015) 07, 1530022.

\bibitem{Wang:2012hu}
  Z.~G.~Wang,
  Commun.\ Theor.\ Phys.\  {\bf 61} (2014) 1,  81.

 \bibitem{Zeynali:2014wya}
  K.~Zeynali, V.~Bashiry and F.~Zolfagharpour,
  Eur.\ Phys.\ J.\ A {\bf 50} (2014) 127.

\bibitem{Bashiry:2014qia}
  V.~Bashiry,
  Adv.\ High Energy Phys.\  {\bf 2014} (2014) 503049.

\bibitem{Xu:2015eev}
  G.~Z.~Xu, Y.~Qiu, C.~P.~Shen and Y.~J.~Zhang,
  arXiv:1601.03386 [hep-ph].

  \bibitem{Chang:2015jla}
  Q.~Chang, P.~P.~Li, X.~H.~Hu and L.~Han,
  Int.\ J.\ Mod.\ Phys.\ A {\bf 30} (2015) no.27,  1550162.

  \bibitem{Chang:2015ead}
  Q.~Chang, X.~Hu, J.~Sun, X.~Wang and Y.~Yang,
  Adv.\ High Energy Phys.\  {\bf 2015} (2015) 767523.

\bibitem{Korner:1987kd}
  J.~G.~Korner and G.~A.~Schuler,
  Z.\ Phys.\ C {\bf 38} (1988) 511 [Erratum: Z.\ Phys.\ C {\bf 41} (1989) 690].

\bibitem{Korner:1989qb}
  J.~G.~Korner and G.~A.~Schuler,
  Z.\ Phys.\ C {\bf 46} (1990) 93.

\bibitem{Hagiwara:1989cu}
  K.~Hagiwara, A.~D.~Martin and M.~F.~Wade,
  Nucl.\ Phys.\ B {\bf 327} (1989) 569.

\bibitem{Hagiwara:1989gza}
  K.~Hagiwara, A.~D.~Martin and M.~F.~Wade,
  Phys.\ Lett.\ B {\bf 228} (1989) 144.

  \bibitem{Kadeer:2005aq}
  A.~Kadeer, J.~G.~Korner and U.~Moosbrugger,
  Eur.\ Phys.\ J.\ C {\bf 59} (2009) 27.


 
\bibitem{Goity:2000dk}
  J.~L.~Goity and W.~Roberts,
  Phys.\ Rev.\ D {\bf 64} (2001) 094007.

\bibitem{Ebert:2002xz}
  D.~Ebert, R.~N.~Faustov and V.~O.~Galkin,
  Phys.\ Lett.\ B {\bf 537} (2002) 241.

\bibitem{Zhu:1996qy}
  S.~L.~Zhu, W.~Y.~P.~Hwang and Z.~s.~Yang,
  Mod.\ Phys.\ Lett.\ A {\bf 12} (1997) 3027.

\bibitem{Aliev:1995wi}
  T.~M.~Aliev, D.~A.~Demir, E.~Iltan and N.~K.~Pak,
  Phys.\ Rev.\ D {\bf 54} (1996) 857.

\bibitem{Colangelo:1993zq}
  P.~Colangelo, F.~De Fazio and G.~Nardulli,
  Phys.\ Lett.\ B {\bf 316} (1993) 555.

  \bibitem{Choi:2007se}
  H.~M.~Choi,
  Phys.\ Rev.\ D {\bf 75} (2007) 073016.

\bibitem{Cheung:2014cka}
  C.~Y.~Cheung and C.~W.~Hwang,
  JHEP {\bf 1404} (2014) 177.


  \bibitem{Wirbel:1985ji}
  M.~Wirbel, B.~Stech and M.~Bauer,
  Z.\ Phys.\ C {\bf 29}  (1985) 637.

\bibitem{Bauer:1988fx}
  M.~Bauer and M.~Wirbel,
  Z.\ Phys.\ C {\bf 42}  (1989) 671.


\bibitem{Caprini:1997mu}
 I.~Caprini, L.~Lellouch and M.~Neubert,
  Nucl.\ Phys.\ B {\bf 530} (1998) 153.

\bibitem{deDivitiis:2008df}
  G.~M.~de Divitiis, R.~Petronzio and N.~Tantalo,
  Nucl.\ Phys.\ B {\bf 807} (2009) 373.
  
\bibitem{LHCb:upgrade}
  LHCb Collaboration,
  CERN-LHCC-2012-007.
  
\bibitem{Buskulic:1995mt}
  D.~Buskulic {\it et al.} [ALEPH Collaboration],
  Z.\ Phys.\ C {\bf 69} (1996) 393.

\bibitem{Bernlochner:2015mya}
  F.~U.~Bernlochner,
  Phys.\ Rev.\ D {\bf 92} (2015) no. 11, 115019.

\bibitem{Tanaka:1994ay}
  M.~Tanaka,
  Z.\ Phys.\ C {\bf 67} (1995) 321.

  
 



\end{thebibliography}
\end{document}